\documentclass[pra,twocolumn,superscriptaddress,showpacs]{revtex4}
\usepackage{graphicx}
\usepackage{natbib}
\usepackage{amssymb} 

\newcommand{\ex}[1]{\langle #1 \rangle}

\newcommand{\ket}[1]{| #1 \rangle}
\newcommand{\bra}[1]{\langle #1 |}

\newcommand{\ew}[1]{\langle #1 \rangle}
\newcommand{\beq}{\begin{eqnarray}}
\newcommand{\eeq}{\end{eqnarray}}

\begin{document}

\title {A macro-realism inequality for opto-electro-mechanical systems}
\author{Neill Lambert}
\affiliation{Advanced Science Institute, RIKEN, Wako-shi, Saitama
351-0198, Japan}
\author{Robert Johansson}
\affiliation{Advanced Science Institute, RIKEN, Wako-shi, Saitama
351-0198, Japan}
\author{Franco Nori}
\affiliation{Advanced Science Institute, RIKEN, Wako-shi, Saitama
351-0198, Japan} \affiliation{Physics Department, University of
Michigan, Ann Arbor, Michigan, 48109, USA} 

\begin{abstract}
We show how to apply the Leggett-Garg inequality to
opto-electro-mechanical systems near their quantum ground state.
We find that by using a dichotomic quantum non-demolition
measurement (via, e.g., an additional circuit-QED measurement
device) either on the cavity or on the nano-mechanical system
itself, the Leggett-Garg inequality is violated.  We argue that
only measurements on the mechanical system itself give a truly
unambigous violation of the Leggett-Garg inequality for the
mechanical system.  In this case, a violation of the Leggett-Garg
inequality indicates physics beyond that of ``macroscopic
realism'' is occurring in the mechanical system. Finally, we
discuss the difficulties in using unbound non-dichotomic
observables with the Leggett-Garg inequality.
\end{abstract}
\maketitle


The Leggett-Garg (LG) inequality~\cite{LG1,LG2,Korot1,Korot2,Vion,
Lambert10,Lambert101,PhotonLG} is one of a large class of
inequalities used to delineate different physical theories.  It is
constructed to test for ``macroscopic realism", the class of
physical theories that imply that before we measure a property of
a system, that property has a well defined value (which is not the
case in quantum mechanics). Bell's inequality~\cite{Bell64} also
tested for this property, but not without also testing for
non-locality. The LG inequality \cite{LG1,LG2} attempts to test
only for realism, but to do so requires the assumption of {\em
non-invasive measurement}, and macroscopically distinct states.
Hence the moniker of ``macroscopic realism".

In the original LG proposal~\cite{LG1}, they imagined measuring
the two different and distinct ``macroscopic states" of a
superconducting flux qubit (where, mathematically, one can
describe these states as a quantum two-level system).  However,
physically these two states are by most definitions
``macroscopic":  They involve millions of particles.
Superconducting qubits have been used to show violations of Bell's
inequality ~\cite{martinis}, the Leggett-Garg inequality
\cite{Vion}, and have been proposed as a way to test the
Kochen-Specker theorem~\cite{KStheorem}.

An alternative candidate to test for quantum behavior in the
macroscopic limit is in the ground state of a nano-mechanical
oscillator~\cite{clelandbook, blencowe}.  Strong evidence has been
reported of success in this goal by coupling a nanomechanical
resonator to a qubit~\cite{oconnell} . Recent work suggests that
the ground state has also been reached in an opto-mechanical
device~\cite{teufel, teufel2}. An opto-mechanical system is
essentially an optical (or microwave) cavity coupled to a
mechanical resonator to cool and measure the mechanical
system\cite{metzger04,gigan06}. A generic physical model for this
opto-mechanical system is of a spring that supports one of the
mirrors of an optical cavity, and thus the mechanical motion of
the spring is coupled to the frequency of the optical mode.
However, the physical realization of opto-mechanical devices can
vary greatly, from a mirror suspended on a
cantilever~\cite{gigan06}, to a mechanical membrane capacitively
coupled to a microwave
transmission line~\cite{teufel,teufel2}.  

Reference~\cite{gigan06} is an interesting example of the {\em
optical}-cavity realization of an opto-mechanical system.
They~\cite{gigan06} showed side-band cooling from photo-pressure,
and evidence of normal-mode splitting, i.e., strong coupling
between optical and mechanical modes.  Recent
results~\cite{teufel, teufel2} using opto-electro-mechanical
systems (i.e., a microwave transmission line in place of the
optical cavity) have shown ultra-strong coupling and ground-state
cooling. However one cannot easily distinguish the resultant
effective low-temperature state of two coupled quantum oscillators
from two coupled classical oscillators \cite{Franco1,Franco2}. It
is well known from quantum optics that the linear response
spectral properties one observes are similar for both
theories~\cite{Walls95, gigan06,clerk10}, though spectral
properties can strongly {\em infer} cooling to the mechanical
ground state~\cite{Marquardt,teufel,teufel2,Franco1}. In addition,
the observation of asymmetry between spectral peaks due to
absorption and emission of quanta is purely a quantum effect
\cite{clerk10}, and has been recently observed in experiment
\cite{painter}.

In this work we propose a method of further distinguishing quantum
and classical oscillators by applying
the Leggett-Garg inequality. 
By using dichotomic 
quantum non-demolition measurements (QND) ~\cite{Johnson,Schuster}
we show that a theoretical model of a realistic opto-mechanical
system implies a violation of the Leggett-Garg inequality due
 to the coherent interaction of the cavity with the mechanical
 oscillator.
 We show how either measurements on the cavity, or on the mechanical
 system directly, produce violations of the inequality.  We argue that the latter
 are stronger proofs of quantum behaviour in the mechanical
 system,
 as the former can also occur due to the quantum nature of the
 cavity alone.  

Since the dichotomic QND measurements we use here require strong
coupling to a qubit our results are most directly applicable to
{\em opto-electro-mechanical} systems which employ microwave
transmission lines (as the ``opto-electro-'' cavity that cools the
mechanical system to its quantum ground state). As far as we are
aware, a {\em dichotomic} number state measurement has not been
achieved in optical cavities.  We believe the results we show here
align well with Leggett and Garg's original goal of testing for
non-realism in the macroscopic world, since the ground state (or a
single Fock state) of a mechanical oscillator represents a quantum
state in a solid composed of millions of atoms.

We begin this article by outlining the original Leggett-Garg
inequality, and discuss why dichotomic QND measurements are
necessary.  We then show our main result: that the introduction of
a single photon into the microwave cavity, and application of {\em
dichotomic QND measurements}, leads to a violation of the LG
inequality. Afterwards we present the technical details of the
model we use to describe the opto-electro-mechanical system,  and
discuss the practical issues of state
preparation and measurement. 
We finish with a discussion of the difficulties of using
non-dichotomic unbound measurements, and give a conjecture on a
possible bound for the inequality in such a case.

\section{The Leggett-Garg inequality}

The Leggett-Garg inequality~\cite{LG1,LG2} is defined as follows:
Given an observable $Q(t)$, which is bound above and
below~\cite{Korot1,Korot2} by $|Q(t)|\leq1$, the assumption of:
(A1) macroscopic realism and (A2) non-invasive measurement
implies,
\beq L(t_1,t_2) &=&
        \ew{Q(t_1) Q(0) }
        +
        \ew{Q(t_1\!+\!t_2)Q(t_1)}\nonumber\\
        &-&
        \ew{Q(t_1\!+\!t_2)Q(0)}
        \leq 1
    \label{LG}
    ,\ \ \
\eeq If $Q$ is in the steady state at the initial time of
measurement, and we set $t_1=t_2$, then this becomes \beq
2\ew{Q(\tau) Q(0) }-\ew{Q(2\tau) Q(0) }\leq 1.\eeq

To adapt this to work on measurements on bosonic (harmonic)
systems one must proceed with extreme caution. This is because (a)
it is difficult to define a bound on measurements on harmonic
systems, (b) many measurements (particularly in the optical
regime) are invasive (e.g., single-photon counting) and (c) the
dynamics of classical and quantum harmonic systems are identical
(apart from quantum fluctuations) without additional sources of
non-linearity.

Fortunately, the growing field of opto-electromechanical systems
and circuit QED~\cite{you032,blais3,you05,you10, Iulia} allows us
to overcome many of these obstacles. We can adapt the scheme
realised by Johnson et al~\cite{Johnson, Schuster} to overcome
obstacle (a). In their scheme, one uses an additional
qubit/measurement-cavity system to dispersively measure whether
the optomechanical-cavity contains ``one photon or not'' (this is
a dichotomic QND measurement). We will also show how, in
principle, it might be possible to use this to measure the
mechanical system directly, and say if it contains ``one phonon or
not''.

The former (a dichotomic QND measurement on the cavity) is
possible due to the strong coupling between qubit and cavity that
has been achieved in circuit-QED systems.  The latter (a
dichotomic QND measurement on the mechancical system) may be
possible given the recent strong coupling shown between a
mechanical system and a superconducting qubit \cite{oconnell}. A
possible realization of the QND measurement on the cavity in an
opto-electro-mechanical system is shown in Fig.~1.

This scheme also allows us to overcome obstacles (b) and (c), as
it realizes a non-demolition, and classically non-invasive,
projective measurement of the photons
in the cavity (or phonons in the mechanical oscillator).  
In the final section we will return to these issues, and
conjecture about a new bound for the inequality if one's
observables are non-dichotomic and unbound.

\begin{figure}[h]
\includegraphics[width=\columnwidth]{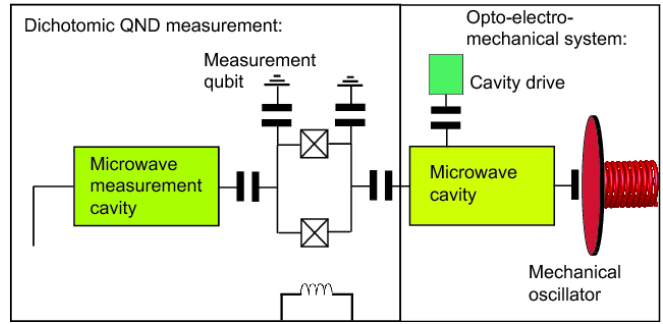}
\caption{{} (Color online)  Schematic diagram of a possible
opto-electromechanical system with dichotomic QND measurement of
the cavity. Right side of the figure: the driven microwave cavity
is coupled to a mechanical oscillator, e.g. the fundamental mode
of a thin film drum as in [\onlinecite{teufel, teufel2}]. The left
side of the
 figure shows a simplified schematic of the qubit/measurement-cavity system used for the dichotomic QND readout (as in Ref.~[\onlinecite{Johnson}]).   A similar measurement could be performed directly on
 the phonon states in the mechanical mode, given sufficiently strong coupling strengths between qubit and mechanical system (such a configuration is not shown in the figure).}
\end{figure}

\begin{figure}[h]
\includegraphics[width=\columnwidth]{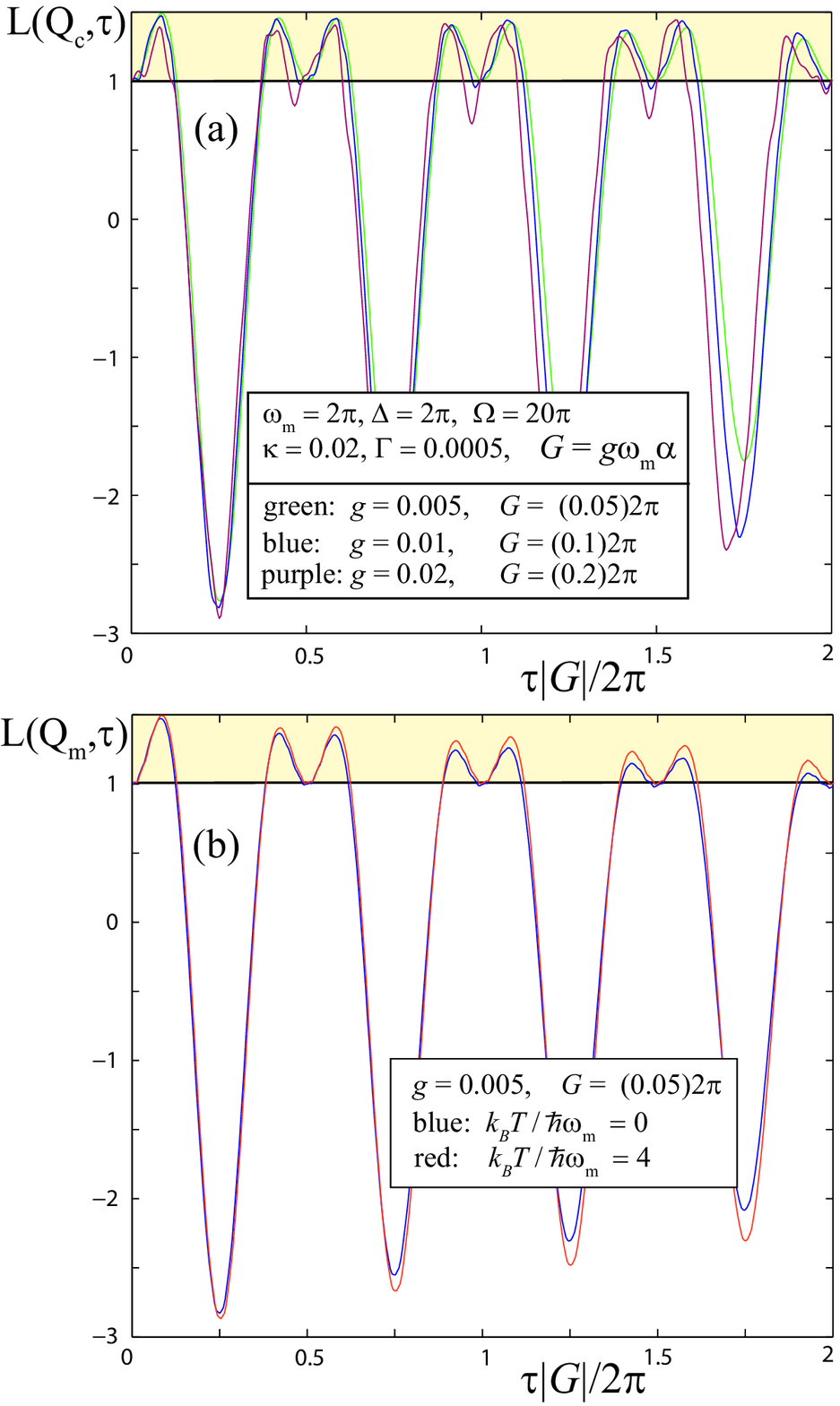}
\caption{{} (Color online). Example of violation of Eq.~(\ref{LG})
(with $t_1=t_2=\tau$) using (a) measurements on the cavity $Q_c$
and (b) measurements on the mechanical mode $Q_m$, as a function
of the dimensionless time: $\tau |G|/2\pi$.  Both figures are
produced by our model of an opto-electromechanical system, where
$G= g\omega_m \alpha$ is the effective coupling between the
mechanical and cavity modes, and $\alpha \approx -\Omega/\Delta$
is the displacement of the cavity mode produced by the driving
$\Omega$. For strong effective coupling, the non-energy conserving
terms in our model begin to strongly modulate the shape of the
correlation functions in Eq.~(\ref{LG}). We have chosen ratios for
the parameters that correspond approximately to those seen in the
strong-driving limit of Ref.~[\onlinecite{teufel,teufel2}], where
$\omega_m \approx 2\pi \times 10$ MHz, though to aid computation
we made $g$ large, and $\Omega$ small, in comparison to their
data.  Remarkably, the high quality factor of both the microwave
and mechanical cavities used in~\cite{teufel,teufel2} means that
the violation of the inequality remains visible for relatively
long time scales, though this can depend on the initial
temperature of the mechanical mode.}
\end{figure}
\section{Violation for optomechanical systems using dichotomic QND measurements}


We define the LG inequality in terms of dichotomic quantum
non-demolition measurements either on the single-Fock state
occupation of the cavity mode  \beq Q_c = 2\ket{1}_c\bra{1}_c -
1,\eeq where $c$ refers to the cavity mode, or on the
single-phonon state occupation of the mechanical mode, \beq Q_m =
2\ket{1}_m\bra{1}_m - 1,\eeq where $m$ refers to the mechanical
mode. As mentioned above these measurements requires an additional
qubit/measurement-cavity~\cite{Johnson}, which we outline in
section \ref{meas}, and is shown schematically in Fig.~1, for the
example of measuring the cavity mode. For our purposes, this
measurement returns $+1$ if there is a quanta in the appropriate
mode ($m$ or $c$), and $-1$ if not. To show a violation of
Eq.~(\ref{LG}) one can prepare the opto-electromechanical system
near its ground state following the side-band cooling procedure
described in the next section. The ground state cooling of the
mechanical system requires that we strongly drive the microwave
cavity, resulting in a non-zero steady-state coherent occupation
in the cavity in a rotating frame. Fortunately this can, in
principle, be
eliminated from the QND measurement (see section \ref{meas}).

We then adiabatically introduce an additional photon into the
cavity~\cite{footnote}, in addition to this coherent
state~\cite{Johnson, milburn}, which ideally prepares the system
in the state \beq \rho(t=0)\ =\  \ket{1}_c\;\bra{1}_c \otimes
\ket{0}_m\;\bra{0}_m,\eeq where again $c$ refers to the cavity and
$m$ refers to the mechanical mode (note that the state of the
cavity is in a displaced basis in a rotating frame because of the
driving of the cavity). The strong coupling between the mechanical
system and the optical mode causes this single excitation to be
coherently exchanged (akin to a Rabi oscillation).  One then
measures the operator $Q_c$ (or $Q_m$) using the readout
qubit-measurement-cavity and a programmable C-NOT
scheme~\cite{Johnson} (see section \ref{meas}). If the measurement
timescale (which includes rapidly resetting the qubit to its
ground state) is short enough one can construct the two-time
correlation functions  in Eq.~(\ref{LG}).

In Fig.~2 we explicitly show how the results from our model,
outlined in the next section, which suggest a violation of
Eq.~(\ref{LG}), is in principle observable with existing
experiments~\cite{teufel,teufel2, oconnell}. It is interesting to
note that the largest violations occur for small times, which
implies we require the readout and reset of the qubit to be fast.
Typical readout times in Ref.~[\onlinecite{Johnson}] are of the
order of $550$ ns, introducing an intrinsic minimum delay into the
correlation functions. The typical time scale of the coherent
dynamics in Ref.~[\onlinecite{teufel,teufel2}] is related to the
coupling, which is of the order of $10^5$--$10^6$ Hz, implying
that the short time-scale coherent dynamics should be observable
with such a measurement, though for measurements of $Q_c$ this may
be altered by the need to measure in a rotating frame (see later).
We also find that if the driving, and hence the coupling between
cavity and mechanical system, is strong enough then the non-energy
conserving terms in the interaction modulate the dynamics quite
strongly. However at this point one also expects other non-linear
affects to arise.

\subsection{Ambiguities in cavity measurements}

As we will discuss shortly, the measurement of $Q_c$ is a direct
adaptation of an existing experiment (albeit with additional steps
to make sure that we are measuring in the correct frame). However,
our main goal is to verify the quantum dynamics of the {\em
mechanical} mode.  It just so happens that in this case it is the
quantum coherent interaction between the cavity and mechanical
modes which drives the violation we observe in the observables of
the cavity system.  However, in principle a violation could also
be observed due to the quantum nature of the cavity mode alone.

Thus with measurements on the cavity mode alone it is impossible
for us to state that a violation the Leggett-Garg inequality
(e.g., with measurements $Q_c$) gives unambiguous proof of
macroscopic quantum phenomena in the mechanical mode.   Ideally,
one  requires dichotomic QND measurements on the mechanical mode
directly (as defined by $Q_m$) to state that a violation of the
Leggett-Garg inequality is unambiguous proof of quantum mechanics
in the nano-{\em mechanical} system. We will outline a possible
scheme to achieve this later.

\section{Optomechanical systems }

We now explicitly describe the optomechanical system, and the
model we use to calculate the results shown in Fig.~1. This model
is well known and studied in other works~\cite{Marquardt, Wilson},
but we provide details here for clarity.

We start with the Hamiltonian describing the coupling between the
cavity and the mechanical oscillator~\cite{Law}, \beq H^{(1)} &=&
\Delta a^{\dagger}a + \omega_m b^{\dagger}b +
g\omega_m(b+b^{\dagger})a^{\dagger}a \nonumber\\&+&
\Omega(a+a^{\dagger}) \label{Hom}\eeq The driving field is such
that the cavity and mechanical mode are now near resonance
($\Delta = \omega_m$).

One of the approaches taken before (e.g., ~\cite{Marquardt,
Wilson}) is to insert displacements $\alpha$ and $\beta$ for both
modes, take the limit $g, g\alpha \ll \Delta,\omega_m$, and treat
the cavity mode as an effective environment which cools the
mechanical mode. The condition for cooling, in our notation, is
then $\Delta
> 0$, and a sufficiently high-quality cavity $\kappa < \omega_m$.  Reference ~\cite{gigan06} has shown that by observing the homodyne
transmission spectra from the (optical) cavity, we can see a clear
signature of ``normal mode splitting".   This is because the
displacements $\alpha$ and $\beta$, and the coupling, are given by
a simple model of two coupled oscillators, which has coupled
normal modes.  Here we explicitly model both cavity and mechanical
system in the strong-coupling regime observed in
Ref.~[\onlinecite{teufel}] using a master equation approach.

\subsection{Resolved side-band cooling}

One can find the linearized version of the original Hamiltonian by
displacing both mechanical and cavity modes, so that $a\rightarrow
c+\alpha$,  $b\rightarrow d+\beta$. Inserting these displacements
in the Hamiltonian, and eliminating linear terms, gives two
coupled equations for the displacements, \beq  \alpha \Delta +
2\omega_m g\alpha (\beta+\beta^*) + \Omega - i\kappa =0 \eeq \beq
\omega_m \beta + \omega_m g\alpha^2 =0 \eeq

The $\kappa$ term arises because of the linear dissipation terms
(cavity losses) we will introduce shortly. Careful inspection of
the possible solutions of these cubic equations shows that, \beq
\beta = -g\alpha\alpha^*,\eeq and in the limit of small $g$,
\beq \alpha \approx \frac{-\Omega}{\Delta - i\kappa/2}. \eeq If we
do not make a small $g$ assumption, these displacements are real,
 up to some critical driving $\Omega$ of order $\omega_m/g$,
which corresponds to the breakdown point of the small displacement
assumption made to derive the original Hamiltonian (see
Ref.~[\onlinecite{Law}]). At this point additional non-linearities
in the interaction could play a role, but we do not consider those
here.

The Hamiltonian, with the linear terms eliminated, becomes \beq
H^{(2)}&=&  (\Delta +2g\omega_m \beta) c^{\dagger}c   + \omega_m d^{\dagger}d\\
&+&g\omega_m (d+d^{\dagger})(\alpha^* c+\alpha c^{\dagger}) +
g\omega_m(d+d^{\dagger})c^{\dagger}c.\nonumber \eeq

We then add standard Lindblad cavity and mechanical losses to this
model, and solve the resulting Master equation, \beq \dot{\rho}
&=& -i[H^{(2)}, \rho]\nonumber \\ &+& \frac{\kappa}{2} \left\{-
c^{\dagger}c \rho - \rho
c^{\dagger}c + 2c\rho c^{\dagger} \right. \nonumber \\
&+& \left. \left[ ( \alpha^*c- \alpha c^{\dagger})\rho + \rho(
\alpha c^{\dagger}- \alpha^* c)\right] \right\}\nonumber \\ &+&
\frac{\Gamma}{2}(\bar{N} +1) \left[-d^{\dagger}d\rho -\rho
d^{\dagger}d
+2d\rho d^{\dagger}\right] \nonumber \\
 &+& \frac{\Gamma}{2}\bar{N} \left[-d d^{\dagger}\rho -\rho dd^{\dagger}
+2 d^{\dagger}\rho d \right] \nonumber \\
&+&\frac{\Gamma}{2}\left[ ( \beta^* d-\beta d^{\dagger})\rho +
\rho(\beta d^{\dagger} -
\beta^* d) \right]\nonumber\\
\eeq Where $\bar{N}$ is the initial thermal occupation of the
mechanical mode. Dissipation terms linear in $c$ and $d$ (and
displacements $\alpha$, $\beta$) arise because of the shifted
coordinate frame.  As mentioned earlier, the linear terms for the
cavity can be easily eliminated by including them in the
displacement $\alpha$. The linear terms for the mechanical mode
dissipation are small in the limit of a high quality factor
resonator, so we neglect them here (though we have numerically
checked that their influence is small).

Under the conditions $\kappa < \omega_m$, $\Delta > 0$ and
sufficiently large driving strength $\Omega$, one can achieve the
well-known resolved side-band-limit cooling; one can start from a
thermal state of the resonator at a given temperature, and reach a
steady-state, where the thermal phonon occupation of the
mechanical system approaches zero. See
Refs.~[\onlinecite{Marquardt, Wilson}] for further details and
discussion of the cooling process.

Using this model we can easily construct the various correlation
functions needed for Eq.~(\ref{LG}). Our technique is to prepare
the system in the appropriate initial state; e.g. a single photon
in the cavity (in the displaced basis) \beq \quad
\rho(0)=\ket{1_c,0_m}\bra{1_c,0_m},\eeq then the appropriate
correlation functions are calculated via the time evolution \beq
\ex{Q_i(2\tau)Q_i(\tau)} = \mathrm{Tr}[Q_i\exp{[L \tau]}
Q_i\exp{[L \tau]} \rho(0)],\eeq or via the quantum regression
theorem.

Since we operate always in the basis of the displaced modes, we
are always close to the steady state. Thus imposing directly as
initial conditions a single Fock state is a sufficiently good
approximation to the true process of preparing the opto-mechanical
system in its steady state, and then, e.g., introducing the
single-photon state using the measurement qubit~\cite{Johnson}. In
principle one can explicitly model this state-preparation
stage~\cite{milburn}, but for simplicity we omit it here.

\section{QND readout}
\label{meas}


As discussed earlier, both of these measurements, $Q_c$ and $Q_m$,
are challenging, but may be feasible in the future by combining
existing circuit-QED devices (for QND readout~\cite{Johnson}) with
an opto-electro-mechanical system~\cite{teufel, teufel2}. The
additional circuit-QED system (qubit and microwave cavity) allows
both the deterministic preparation of the cavity in a single Fock
state~\cite{LiuFock, Fockstates, Schuster,  weig}, and the
dispersive QND readout of its population dynamics~\cite{Johnson}.
Thus in reality our proposed opto-electro-mechanical system is a
circuit-QED-mechanical system, where the additional qubit-cavity
part is used for state preparation and readout.  First of all we
will describe the details (Fig. 1) of how to realize the
measurement $Q_c$ (Ref.~[\onlinecite{Johnson}]).  As mentioned
earlier, this is perhaps the most feasible with current
technology, though does not give us an unambigous violation of the
LG inequality for the mechanical mode. Then we will discuss
possible ways that the measurement $Q_m$ might be realised by
coupling the qubit directly to the mechanical system, and not the
cavity, which give us a more ideal and unambigous violation of the
LG inequality.

\subsection{Cavity measurement $Q_c$}

The dichotomic QND measurement realized by Johnson et al
[\onlinecite{Johnson}] is ideal for our purposes of realizing
$Q_c$, but the scheme as it is described there~\cite{Johnson}
measures ``if there is one cavity or not'' in the lab basis.  The
cavity in the opto-mechanical Hamiltonian Eq.~(\ref{Hom}) we used
earlier is in a displaced rotating frame (because of the microwave
driving needed for sideband cooling), and thus it is in this basis
that we must measure the cavity to realize $Q_c$ as we have
described it. For example, in the stationary frame of the qubit
(but rotating frame of the cavity), the interaction between qubit
and cavity is described by,
 \beq H_{qb-c}
 &=& \frac{\epsilon}{2} \sigma_z  + \Delta a^{\dagger}a +
 \lambda\left(\sigma_+ ae^{-i\omega_d t} + \sigma_- a^{\dagger}e^{i\omega_d
 t}\right),\nonumber\\
 \eeq
where $\omega_d$ is the driving frequency, and was chosen to bring
the cavity and mechanical system on resonance in Eq. (\ref{Hom}),
so that \beq \Delta = \omega_c - \omega_d \approx \omega_m.\eeq
In addition, we also displace the cavity co-ordinates by $\alpha$,
so that the interaction between the qubit and cavity co-ordinates
that we actually want to measure is
 \beq \bar{H}_{qb-c}
 &=& \frac{\epsilon}{2} \sigma_z  + \Delta c^{\dagger}c +
 \lambda\left(\sigma_+ c e^{-i\omega_d t} + \sigma_- c^{\dagger} e^{i\omega_d
 t}\right) \nonumber \\ &+& \lambda\left(\alpha e^{-i\omega_d t}\sigma_+  + \alpha^* e^{i\omega_d t} \sigma_-\right),
 \eeq
 The additional displacement term represents the large number of
 photons that are in the cavity due to the driving.  Ideally their
 influence on the qubit can be eliminated by applying an
 additional microwave drive to the qubit itself~\cite{blais3} (still in the lab frame), out of phase with
 the term above, e.g., \beq H_{\mathrm{adjust}} =  -\lambda\left(\alpha e^{-i\omega_d t}\sigma_+  + \alpha^* e^{i\omega_d
 t}
 \sigma_-\right).\eeq    This is feasible if the magnitude, $\lambda \alpha$, is not too
 large~\cite{blais3}, but may become unfeasible if an extremely large driving of
 the cavity is needed for cooling.

Assuming this term has been applied, and the effect of the large
cavity population eliminated, we can move the qubit into the same
frame as the cavity with the unitary transformation
$U_q=\exp\left(i \omega_d \sigma_z t/2\right)$.  This leaves us
with a normal Jaynes-Cummings Hamiltonian between qubit and
cavity, with a shifted qubit energy \beq \Delta' = \epsilon -
\omega_d.\eeq

 In this new picture, the QND measurement scheme proposed and analyzed elsewhere~\cite{blais3, Clerk, Johnson} applies for a large
 bias \beq \delta  = \Delta' - \Delta=\epsilon - \omega_c.\eeq  This is clearly shown by applying the unitary transformation
 $U_dHU_d^{\dagger}$, $U_d = \exp{\left[\frac{\lambda}{\delta}(c \sigma_+ -
 c^{\dagger}\sigma_-)\right]}$,  which leads to the well-known dispersive
 coupling Hamiltonian (see, e.g., Refs.~[\onlinecite{blais3, Clerk}]),
 \beq H_{qb-D}
 &=& \left[\frac{\Delta'}{2}+ \frac{\lambda^2}{\delta}\left(c^{\dagger}c +\frac{1}{2}\right)\right]  \sigma_z
+\Delta c^{\dagger}c.
 \eeq
This transformation can induce interactions between the
measurement qubit and the mechanical mode, but these terms can
also be treated with a dispersive transformation, and give a shift
of the qubit frequency of order $(\lambda g \alpha
\omega_m)^2/\delta^3$, and are thus much weaker than the
$\lambda^2/\delta$ shift.  In addition, the higher-order terms in
$\lambda/\delta$ (representing back-action of the qubit on the
cavity) should be much smaller than the cavity-mechanical mode
interaction (i.e., $\lambda^3/\delta^2 \ll g \alpha \omega_m$).

In Ref.~[\onlinecite{Johnson}], in order to have sufficiently high
resolution measurement of the effect of the photons on the energy
levels of the qubit, they needed $\delta/\lambda < 10$. That is,
they need sufficiently large $\delta$ to reach the dispersive
limit, but sufficiently strong $\lambda$ to obtain well resolved
energy shifts for different photon occupations. Here, $\delta =
 = \epsilon - \omega_c$, thus reaching the same regime as $\delta/\lambda < 10$ seems
feasible.

The dichotomic property of the measurement is achieved because of
the strong dependence of the qubit response on the number of
photons in the cavity.  In Ref.~[\onlinecite{Johnson}], for the
measurement step, they apply a $\pi$ control pulse to the qubit at
the frequency corresponding to its energy when just one photon is
in the cavity.  Thus the qubit is rotated if and only if there is
one photon present, and nothing happens otherwise. This is an
effective CNOT gate on the qubit and the cavity.  Here, the
effective CNOT gate must also be in the rotating frame of the
qubit. For the final measurement step, one measures the state of
the qubit via pulsed spectroscopy of the cavity.

Overall, this dispersive Hamiltonian, combined with the
controlled-$\pi$-rotation of the qubit, and readout of the qubit
using the additional measurement cavity (which we have not
explicitly described), ideally gives us a way to realize the
dichotomic QND measurement $Q_c=2\ket{1}_c\bra{1}_c -1$. As we
discussed earlier, the time needed in Ref.~[\onlinecite{Johnson}]
to realize this measurement may be short enough to observe
correlation functions on the time scale we require.  However, in
general there will be losses involved in the measurement process
(e.g., due to dissipation of the qubit state) which will degrade
the measurement result~\cite{Johnson}. In addition, the need to
perform the effective CNOT gate in the rotating frame may slow
down the measurement step~\cite{blais3}.

\subsection{Mechanical measurement $Q_m$}

As we have reiterated several times, a measurement of $Q_c$ is not
sufficient to {\em unambiguously} show quantum dynamics in the
mechanical mode.  We ideally need to perform the dichotomic QND
measurement on the mechanical system itself.  As far as we are
aware no similar measurement has yet been achieved, though efforts
on membrane-in-the-middle devices \cite{clerk10} are promising.
Staying within the regime of the nano-mechanical systems we have
discussed so far \cite{teufel, teufel2, oconnell}, one can imagine
adapting the scheme of Johnson et al [\onlinecite{Johnson}] to
directly measure the mechanical mode. In
Ref.~[\onlinecite{oconnell}] O'Connell et al observed a strong
interaction between a high-frequency mechanical mode and a
superconducting qubit. There the mechanical mode frequency was
$\omega_m/2\pi = 6$ GHz, so cryogenic freezing was sufficient to
reach the quantum ground state, and they observed qubit-oscillator
coupling strengths of $\lambda/h = 110$ MHz.  This is favorable
for using the qubit as a dispersive measurement of the mechanical
mode. However a straight adaptation of Ref.~[\onlinecite{Johnson}]
to the system in Ref.~[\onlinecite{oconnell}] would have to
compensate for the extremely short quality factor of the
mechanical resonator (the mechanical dephasing time is estimated
to be $T\approx 20$ ns). This is well short of the measurement
time in Ref.~[\onlinecite{Johnson}], and thus resolving the short
time correlations needed to see a violation of the inequality may
prove difficult without improvements in the readout and reset
times of the qubit, or employing a higher quality factor/lower
frequency mechanical resonator.

Furthermore,  one can imagine a similar scenario using the
opto-mechanical side-band cooling systems we have described here,
where the low frequency (and high quality factor) mechanical
system is cooled by the cavity, and then the mechanical part is
measured in the same manner as above (by an additional
superconducting qubit, with compensation for the coherent
occupation $\beta$). This is quite a speculative scenario, as it
is not clear if a sufficiently large coupling between the qubit
and the type of mechanical oscillator used in
opto-electro-mechanical devices ~\cite{teufel,teufel2} can be
engineered, and if the overly large energy mismatch between the
superconducting qubit and mechanical resonator overcome. However
if realised it would be ideal for truly showing macroscopic
quantum phenomena in the same spirit as Leggett and Garg's
proposal \cite{LG1}.

\subsection{Single photon measurements}

For the optical-cavity realization of an optomechancial system (as
in Ref.~\cite{gigan06}),  the Fock state preparation (e.g., by
using an additional one-way cavity as discussed in
Ref.~[\onlinecite{milburn}]), and QND measurements~\cite{Guerlin},
are feasible but the dichotomic measurements we require are much
more difficult to realise than in the microwave cavity case.

We also point out that the correlators in Eq.~(\ref{LG}) are not
normal ordered, and thus do not represent the measurements
obtained from single-photon counting (which are typical in optical
cavity systems). As we discussed in earlier
work~\cite{Lambert10,Lambert101}, photon absorbtion measurements
are fundamentally invasive, and typically represent an obstacle
for un-conditionally verifying quantum behavior via the
Leggett-Garg inequality. This is particularly true with a fragile
single-quantum Fock state, hence the need for QND measurements.

\section{Non-dichotomic and unbound observables}

What happens if we attempt to construct the Leggett-Garg
inequality from non-dichotomic and unbound observables?  Recent
work on Bell's inequality with unbound measurements~\cite{belzig1,
belzig2} suggest that one has to move to fourth-order correlation
functions to distinguish quantum and classical correlations, which
may also apply to the Leggett-Garg inequality.

First of all, let us consider a general picture where we measure
an unbound operator $\ex{\hat{Q}} \in \{-\infty, \infty\}$.
Following the same reasoning as used in the Leggett-Garg
inequality one can derive a bound (and assuming we construct our
expectation values by counting how often a particular measurement
result arises), \beq L_Q\leq\ex{\mathrm{Max}_t[Q(t)]^2}.\eeq Such
a bound may occur due to some intrinsic conservation rule in the
system (e.g., if the number state or energy is conserved).
However, this bound is both difficult to calculate (and measure)
and is extremely loose, since in general extremum values may be
observed but contribute little to the expectation values. Since
the maximization is a convex function, we know that \beq
\mathrm{Max}_t[\ex{Q(t)^2}]\leq\ex{\mathrm{Max}_t[Q(t)]^2},\eeq
(this is the Jensen inequality), but finding further constraints
on these functions is challenging beyond trivial cases.  We
conjecture that there might be a tighter bound for the inequality
given by $\mathrm{Max}_t[\ex{Q(t)^2}]$. 
However we have been unable to find a rigorous proof, and it may
be that a simple counter-example exists to show that this
conjecture does not hold.

There is a further caveat on such an approach. Note that taking
the stationary limit and setting ($t_1=t_2=\tau$) simplifies the
inequality  with our conjectured bound
$\mathrm{Max}_t[\ex{Q(t)^2}]$ to, $ L_{Q}^{(2)}=2\ew{Q(\tau)Q(0)}
- \ew{Q(2\tau)Q(0)}\leq \ex{Q(0)^2}.$ In a classical situation,
the correlation functions one can observe are harmonic functions
(even in the stationary state). For example, one can easily solve
the equation of motion for a single oscillator in contact with a
thermal bath, and find that the spectral density of the
displacement~\cite{clelandbook} is a Lorentzian with amplitude
$\frac{2k_BT}{\pi m Q}$ related to the bath temperature $T$. 
The Wiener-Khinchin theorem tells us that the spectral density is
the Fourier transform of the auto-correlation function in the
steady state, implying sinusoidal correlation functions which are
only dependent on one time variable (the time between
measurements). These obviously cause a violation of the bound
$\ex{Q(0)^2}$ in the steady state.  One can argue that this is not
{\it per se} a failure of our conjectured bound, but is because
the only output one observes from the system is noise-driven; the
thermal background has a white noise spectrum, which can thus
excite the system around its resonant frequency. Thus, in the
language of Leggett-Garg, these {\em classical} correlation
functions are essentially invasive since one only observes a
signal when the system fluctuates (e.g., due to thermal
fluctuations). Therefore, the observation of thermal-noise-induced
fluctuations is
equivalent to a perturbation of the system by the measurement.  

  The original Leggett-Garg
inequality avoids this problem by demanding that the system must
be in one of two macroscopically distinct states, and that the
system is almost
always in one of the two states~\cite{LG1,LG2}.  
  In a harmonic system this assumption,
of macroscopically distinct states, breaks down spectacularly. One
can overcome this problem by avoiding the steady state, or
introducing a third measurement (if $Q \in \{0,\infty\}$, or a
third and fourth measurement if $Q \in \{-\infty, \infty\}$) into
all correlation functions in the inequality at time $t=0$, and
scaling the bound appropriately.
 Then the violation is dependent on the effect of the second
measurement (which we assume again to be non-invasive). The
quasi-invasive (fluctuation) nature of the first measurement
becomes irrelevant.  Introducing extra measurements into the
inequality is akin to the fourth-order Bell inequality derived by
Bednorz et al~\cite{belzig1,belzig2}.  However more work remains
to be done to derive rigorous proofs for our conjecture.

\section{Conclusions}

In summary, we discussed how to use the Leggett-Garg inequality to
distinguish quantum and classical dynamics in
opto-electro-mechanical systems.  We illustrated that dichotomic
QND measurements of either the cavity or mechanical system leads
to a violation of this inequality.  We discussed possible methods
to realize such measurements, and argued that only measurements
directly on the mechanical system itself will give unambigous
proof of macroscopic quantum dynamics in the mechanical system
(or, in the language of Leggett and Garg, proof of a violation of
macroscopic realism).

\section{Acknowledgments}
We thank T. Brandes, C. Emary, S. De Liberato and I. Mahboob for
useful discussions.  NL is supported by the RIKEN FPR program. JR
is supported the Japanese Society for the Promotion of Science
(JSPS). FN acknowledges partial support from the Laboratory of
Physical Sciences, National Security Agency, Army Research Office,
Defense Advanced Research Projects Agency, Air Force Office of
Scientific Research, National Science Foundation Grant No.
0726909, JSPS-RFBR Contract No. 09-02-92114, Grant-in-Aid for
Scientific Research (S), MEXT Kakenhi on Quantum Cybernetics, and
the JSPS via its FIRST program.

\bibliography{bibliography}

\begin{thebibliography}{42}
\bibitem[{foo()}]{footnote}
\bibinfo{journal}{Actually, the steady state is also sufficient as an initial
  state, since $\ex{Q^2}=1$. However for conceptual reasons we use a single
  photon as the initial state here. In addition, if one uses the steady state
  as the initial state, and it is very close to the pure ground state, then the
  violation of the inequality becomes difficult to observe.}
\end{thebibliography}

\end{document}